# RT-utils: A Minimal Python Library for RT-struct Manipulation

Asim Shrestha[1], Adam Watkins[1], Fereshteh Yousefirizi[1], Arman Rahmim[1,2,3,4], Carlos F. Uribe[1,2,5]

[1] Department of Integrative Oncology, BC Cancer Research Institute, Vancouver, British Columbia, Canada
[2] Department of Radiology, University of British Columbia, Vancouver, BC, Canada
[3] Departments of Physics and Biomedical Engineering, University of British Columbia, Vancouver, BC, Canada
[4] Department of Biomedical Engineering, University of British Columbia, Vancouver, Canada
[5] BC Cancer, Vancouver, Canada

**Abstract** – Towards the need for automated and precise AI-based analysis of medical images, we present RT-utils, a specialized Python library tuned for the manipulation of radiotherapy (RT) structures stored in DICOM format. RT-utils excels in converting the polygon contours into binary masks, ensuring accuracy and efficiency. By converting DICOM RT structures into standardized formats such as NumPy arrays and SimpleITK Images, RT-utils optimizes inputs for computational solutions such as AI-based automated segmentation techniques or radiomics analysis. Since its inception in 2020, RT-utils has been used extensively with a focus on simplifying complex data processing tasks. RT-utils offers researchers a powerful solution to enhance workflows and drive significant advancements in medical imaging.

## Introduction

The growing need for automated and robust analysis of medical images (e.g. detection, segmentation, outcome prediction, etc.) has prompted increasingly exploration and adoption of AI-based approaches. These methods often rely on Digital Imaging and Communications in Medicine (DICOM) images (https://www.dicomstandard.org ) and utilize radiation therapy (RT) structures as masks. However, the efficacy of AI approaches can vary across different data sources or conversion techniques [1–3]. To address this challenge, the DICOM standard includes the "radiotherapy structure set (RT-Struct)" object, designed for transferring patient structures and associated information among devices within and outside the radiotherapy department. This object primarily encompasses data related to regions of interest (ROIs) and points of interest, including dose reference points.

While there are available tools for converting DICOM images and RT-Structures into alternative data formats [3,4], the practical integration of auto-segmentation solutions using deep learning techniques in clinical environments remains infrequent due to the lack of accessible open-source frameworks proficient in handling DICOM RT-Structure sets. Additionally, the scarcity of specialized Python modules for converting NumPy arrays into RT-Structures further contributes to this challenge. To address the need for the conversion between discrete medical images and DICOM RT-Structs formats, a number of software packages such as dcmrtstruct2nii (https://github.com/Sikerdebaard/dcmrtstruct2nii), DicomRTTool [4] and PyRaDiSe [3] provide essential functionalities. In the domain of deep learning model inference and image transformations, TorchIO [5] and MONAI (Medical Open Network for AI) [6] are also examples of software packages that provide significant value. Although MONAI is specifically tailored for clinical workflows, both TorchIO and MONAI currently encounter limitations in effectively processing DICOM RT-structure data [3].

In a multi-center investigation [1], the impact of employing diverse mask generation methods on clustering the patients was assessed in the context of radiomics. It was demonstrated that despite utilizing the same dataset, variations in masks and features emerge based on the specific contour-to-mask conversion techniques employed by different medical imaging software. This variability influences patient clustering and has potential implications for radiomic-based modeling in multi-center studies utilizing a combination of mask generation software. In a recent study [2], we studied the reproducibility problem in AI research through code sharing, with a particular emphasis on inconsistent pre- and/or post-processing steps can lead to divergent results. Our findings highlighted the significance of preprocessing steps including the DICOM to NIfTI conversion and RT-struct to mask conversions tools/software and available codes.

Considering the need for increased efficiency and time-saving in the preprocessing and prediction stages of medical research, we have created a specialized Python library, RT-utils (https://github.com/qurit/rt-utils/). This tool is designed to enhance the efficiency of manipulating RT-Structures, offering a comprehensive solution for researchers involved in medical imaging data. To tackle the inherent challenges in data processing within medical research, by developing this library we aimed to optimize workflows and make substantial contributions to advancements in the field. The focus is on simplifying and refining the intricate processes involved in working with medical imaging data, providing researchers with a powerful tool for effective data manipulation.

In this paper, we introduce the robust framework of RT-utils designed for the efficient curation of RT-struct files, employing advanced techniques to contours provided by expert and to generate the output of AI tools output masks to RT-struct format to be used in the clinical workflow.

## Methods

Our module introduces intuitive techniques for efficient data curation of RT-Structure files, facilitating the identification of distinct region of interest (ROI) names and their corresponding locations within the structures. It adeptly handles scenarios where multiple ROI names correspond to the same structure, ensuring a comprehensive and accurate representation. Additionally, the module offers the conversion of DICOM images and RT-Struct into widely used formats such as NumPy arrays and SimpleITK Images. These standardized formats serve as optimal inputs for various applications, including deep learning models, image analysis, and radiomic feature calculations (extraction). Moreover, the toolkit simplifies the process of creating DICOM RT-Struct from predicted NumPy arrays, commonly the outputs of semantic segmentation deep learning models, providing a versatile solution for researchers and practitioners in medical imaging.

In the realm of data science, discretized image formats such as NIfTI, NRRD, and MHA are commonly employed, while radiotherapy workflows heavily rely on the DICOM format, specifically the DICOM RT-Struct. Unlike data science architectures like U-Net, which operate on grid-based data, handling the continuously spaced contour points present in RT-Struct poses a unique challenge. To bridge this gap, accurate data conversion between discrete and continuous spaces becomes crucial when working with clinical DICOM RT-Struct data.

**Technical Overview**

The module stands out in its capability to identify discrete Regions of Interest (ROI) and pinpoint their precise spatial locations within intricate structures. Noteworthy is its adept handling of scenarios where multiple ROI names correspond to the same anatomical structure, ensuring a comprehensive and accurate representation of data.

One of the primary functionalities of the module involves the sophisticated identification of distinct ROI names within RT-Struct files. It also accommodates cases where multiple ROI names map to the same anatomical structure, ensuring precise and comprehensive data organization. The module further facilitates the conversion of DICOM images and RT-Structures into widely adopted formats, specifically NumPy arrays and SimpleITK Images. These standardized formats serve as optimal inputs for various applications, including deep learning models, image analysis pipelines, and radiomic feature calculations.

Another noteworthy feature is the module ability to simplify the generation of DICOM RT-Struct from predicted NumPy arrays, commonly serving as the output of semantic segmentation deep learning models. This functionality establishes a direct link between predictive modeling and clinical applications, offering a path from automated segmentation to structured RT data. RT-utils employs geometric operations from the Shapely library to convert DICOM ROI Contours into binary masks. This involves creating Shapely polygons from contours and rasterizing these polygons into binary masks. Leveraging geometric principles and libraries, our method accurately represents contours as masks, offering a valuable tool for efficient processing of RT-Struct data in radiotherapy applications.

**Practical Applications**

First released in 2020, RT-utils is openly hosted on GitHub (https://github.com/qurit/rt-utils/) (*Starred >160 times as of February 2024), and on PyPI (https://pypi.org/project/rt-utils/) encouraging collaborative development. The demonstrated utility of RT-utils aligns with the evolving landscape of AI-based approaches in the medical domain. In this section, we elaborate on the practical applications of RT-utils. To install RT-utils, simply execute the command `pip install RT-utils`.

*Available Manipulations*

Once installed, users can import `RTStructBuilder` to create a new RT-Struct or load an existing one. Upon accomplishing this, users acquire the capability to execute the following operations:

- o Create a distinctive ROI within the RT-Struct using a binary mask. Optionally, users can specify the color, name, and description of the RT-Struct.
- o Retrieve a list of names for all ROIs contained within the RT-Struct.
- o Load ROI masks from the RT-Struct by specifying their respective names.
- o Safeguard the RT-Struct by providing either a name or a path for storage.

*Handling of DICOM Headers*

RT-utils approach to managing DICOM headers is straightforward and it is designed for simplicity and effectiveness. Initially, we include all necessary headers to ensure the RT-Struct file validity. Subsequently, we maximize the transfer of headers from the input DICOM series, encompassing vital patient information. Moreover, the introduction of new ROIs dynamically integrates them into the relevant sequences within the RT-Struct.

*Incorporating an ROI Mask*

The `add_roi` method requires the ROI mask to follow a specific format: it should be a 3D array where each 2D slice constitutes a binary mask. In this context, a pixel value of one within the mask indicates that the pixel belongs to the ROI for a specific slice. It is essential that the number of slices in the array match the number of slices in your DICOM series. The order of slices in the array should align with their corresponding slice locations in the DICOM series, ensuring that the first slice in the array corresponds to the smallest slice location in the DICOM series, the second slice corresponds to the second slice location, and so on.

## Results

For comparing the effects of different RT-Struct conversion methods, we investigated the RT-utils tool, dcmrtstruct2nii (https://github.com/Sikerdebaard/dcmrtstruct2nii) and the built-in tools from LIFEx [7] and 3D Slicer [8]. We implemented the conversion technique and conducted a comparison of the NIfTI ground truth files. The level of agreement observed between RT-utils and LIFEx surpasses that of other techniques. The mean absolute errors with respect to RT-utils are shown on sagittal and coronal masks. (**Figures 1**). The visual inspection of an example of converted masks overlaid on PET scans using different techniques are shown in **Figures 2** and **3**.

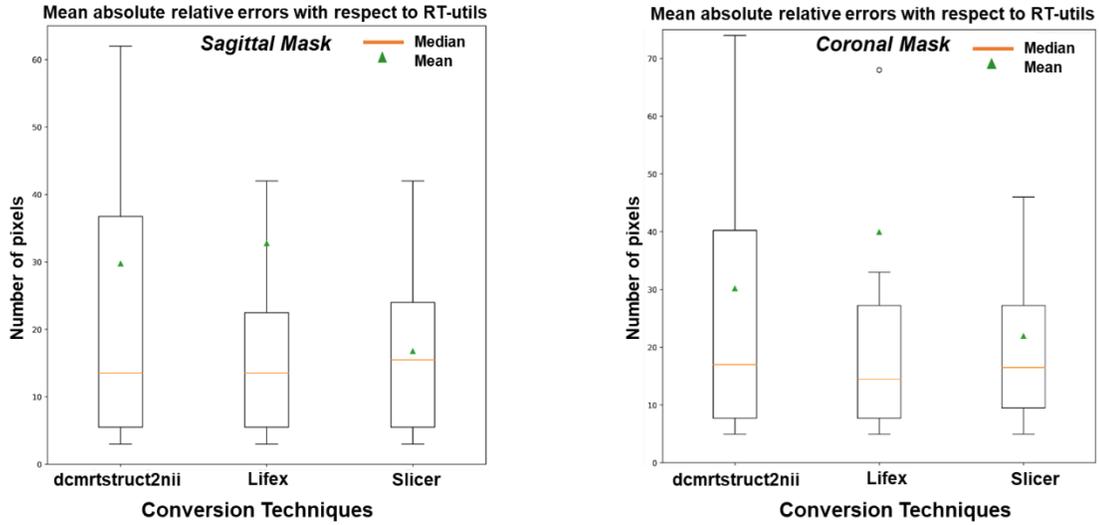

**Figure 1.** Ground truth Mask conversion effects. The mean absolute error with respect to RT-utils are shown on sagittal and coronal masks. The agreement between Rt-util and LIFEx are higher compared to other techniques

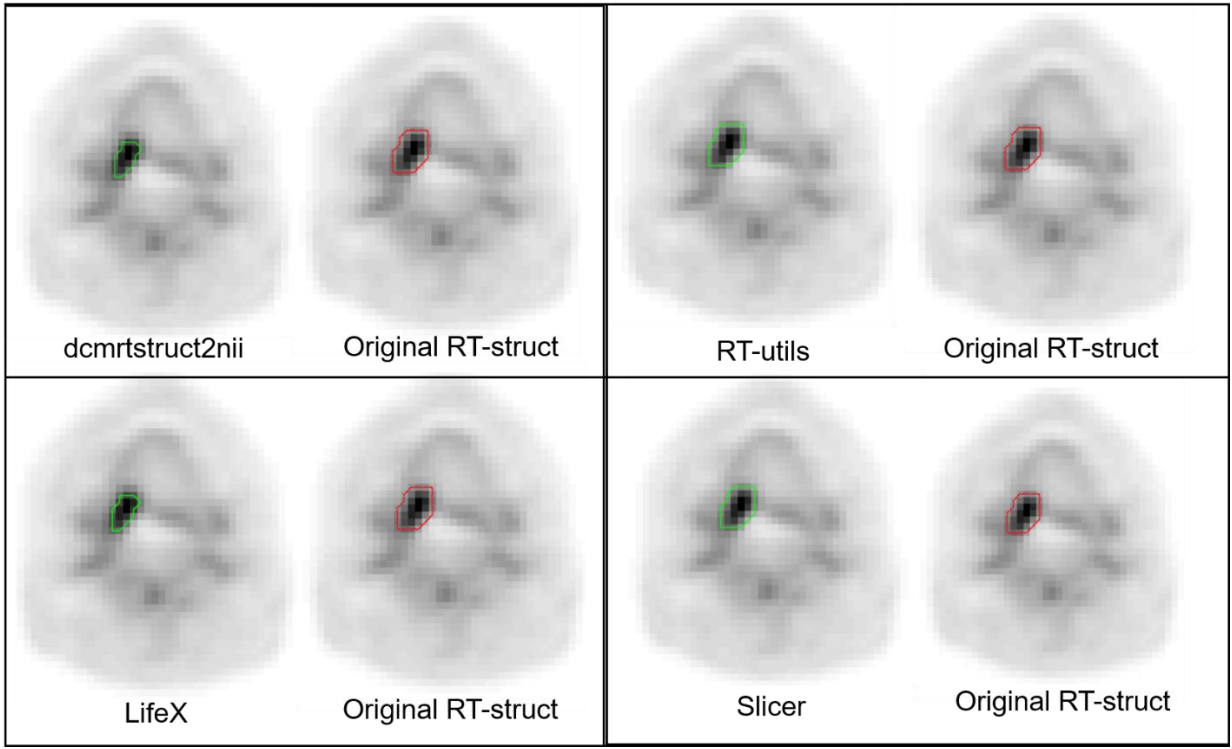

**Figure 2.** The axial view of PET scans is shown by overlaying masks generated using various conversion techniques. While Slicer and RT-util maintain the original size of the RT-struct contour, dcmrtstruct2nii and LIFEx may alter its dimensions.

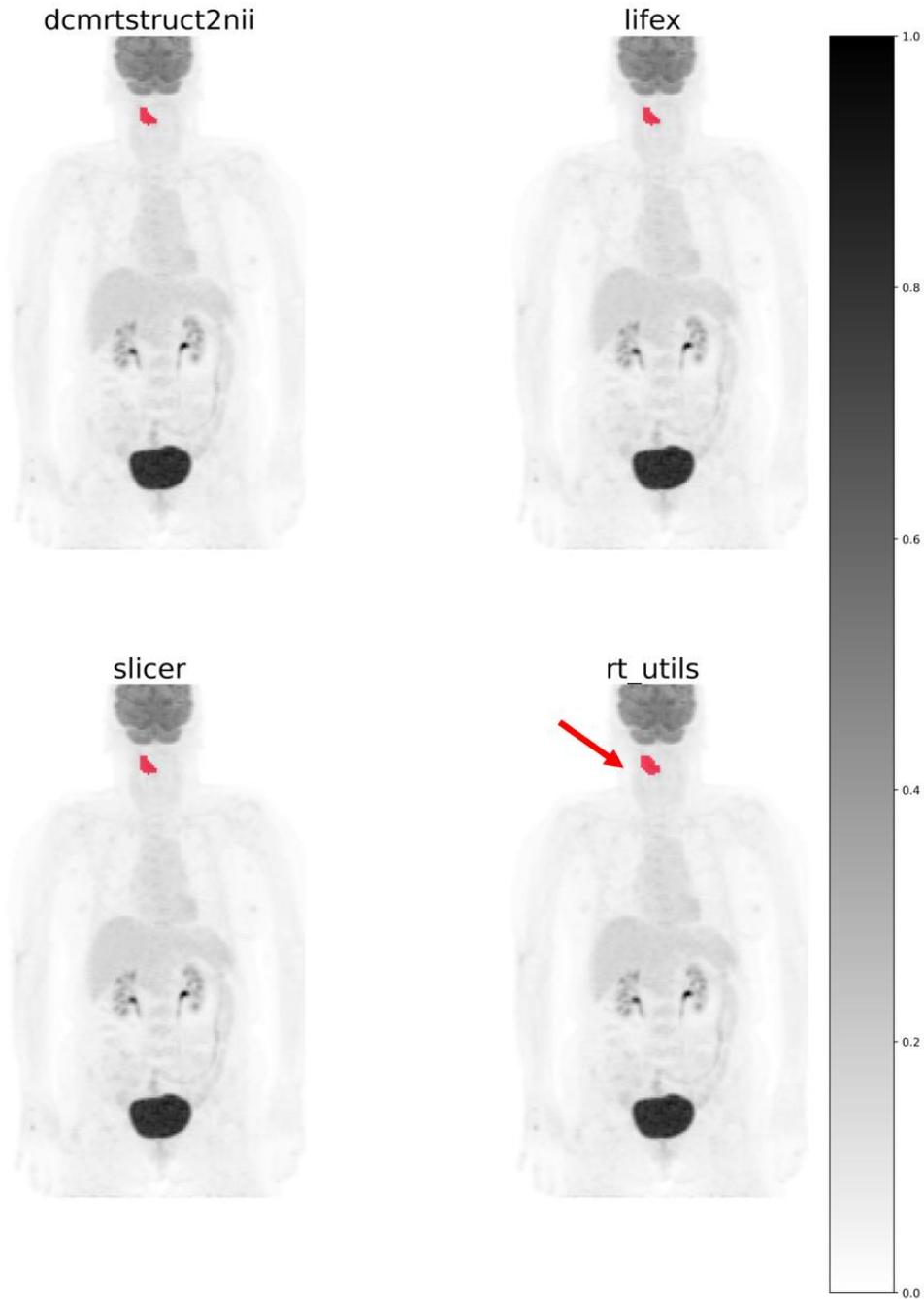

**Figure 3.** RT-Struct to Mask conversions. Converted masks are overlaid on PET scans and they are visually different. RT-utils: Our approach. PET SUV values are scaled in the range of (0,1).

## Discussion and Conclusion

Apart from the data format challenges, developing auto-segmentation solutions involves accurate pre-processing and post-processing steps for DICOM image series. Mask discrepancy would affect the supervised AI based approaches including segmentation [2] and radiomics analyses [10]. Despite existing efforts, a comprehensive Python package to address the diverse challenges of converting RT-structs to ground truth masks and converting the predicted mask by AI model to RT-structs is yet to be realized. We developed a Python package that comprehensively addresses the complexities associated with developing radiotherapy-oriented, DL-based auto-segmentation solutions for clinical DICOM data. By RT-utils, we aimed to bridge the gap between data science and clinical applications by providing a versatile DL inference framework for medical applications.

RT-utils spans a diverse range of technical capabilities such as Creating new RT Structs, Adding to existing RT Structs, loading an existing RT Struct contour as a mask and Merging two existing RT Structs. Rt-utils also has the parameter `use_pin_hole` is a Boolean value that is initially set to false. When enabled (set to true), it erases lines within a mask, allowing each distinct region in an image to be enclosed by a single contour instead of having nested contours. This feature is useful when working with RT-Struct viewers that do not support nested contours or contours with holes. These capabilities extend to various applications, offering accelerated development of deep learning models through standardized inputs. It facilitates the integration of RT-Struct data into computational analyses and image processing pipelines (e.g. radiomics and AI), contributing to the efficiency of medical image analysis. Moreover, the toolkit supports a smooth transition from predictive models to clinical workflows, enhancing the practical utility of automated segmentation. In essence, RT-utils not only simplifies the curation of RT-Struct data but also empowers users with versatile tools for interfacing with standard formats, thereby facilitating advanced medical image analysis and model integration. RT-utils is confined to a straightforward 2D-based conversion algorithm. This limitation might generate a synthetic appearance of RT contour i.e. pixelated contours which could potentially impede the acceptance of generated RT contours within clinical environments and in our future efforts this issue will be addressed.

**Acknowledgements**

The authors wish to acknowledge the Natural Sciences and Engineering Research Council of Canada (NSERC) Discovery Grants RGPIN-2019-06467 and RGPIN-2021-02965.